\documentclass{article}
\usepackage{latexsym, amssymb, amsmath}

     \def\dl{\delta}           \def\sq{\sqrt} \def\fr{\frac} \def\half{\frac{1}{2}}

\def\lap3{~| \!\!\! \partial^2} \def\dlap3{~| \!\!\! \partial^4} \def\invlap3{~| \!\!\! \partial^{-2}}
\def\lang{\langle} \def\rang{\rangle}

\begin{document}

\begin{center}
{\large {\bf Ostrogradsky's Theorem is Incompatible with Background Independence in Quantum Gravity}}
\end{center}

\begin{center}
{\sc Ken-ji Hamada}
\end{center}

\begin{center}
{\it Institute of Particle and Nuclear Studies, KEK, Tsukuba 305-0801, Japan  \\ and
Department of Particle and Nuclear Physics, The Graduate University for Advanced Studies (SOKENDAI), Tsukuba 305-0801, Japan}
\end{center}

\begin{abstract}
Ostrogradsky's theorem shows that higher-derivative dynamical systems inevitably yields a Hamiltonian featuring a ghost mode that is unbounded from below. However, we argue that this theorem is inapplicable to gravitational systems, as the Hamiltonian constraint, which encodes background independence by dictating that the total Hamiltonian vanishes, holds strictly. The ghost mode obeying this constraint constitutes an indispensable component in the construction of space and time. 

Nevertheless, it should be underscored that the use of the weak-field (graviton) approximation, being a background-dependent scheme presupposing the existence of absolute time, invokes this theorem; consequently, it is allowed only in the domain below the Planck scale, where second derivative terms dominate.
\end{abstract}

\section{Introduction and summary}

Adopting diffeomorphism invariance and renormalizability as guiding principles in quantum gravity yields a Schwinger-Dyson equation that enforces the strict vanishing of the expectation value of the total energy-momentum tensor. This entails establishing physical states in which the total Hamiltonian is identically zero. Consequently, Ostrogradsky's classic theorem \cite{ostrogradsky} which asserts the unboundedness of the Hamiltonian for higher-derivative systems is fundamentally inapplicable to this framework of quantum gravity.

However, if we attempt to formulate quantum gravity as an extension of Lorentz-invariant systems via the conventional weak-field approximation, this theorem inevitably comes into play. Therefore, the validity of the weak-field approximation is restricted to gravitational theories featuring at most second derivatives in their kinetic terms, or to domains where such terms constitute the dominant contribution. Accordingly, this approach  cannot be extrapolated beyond the Planck scale. To quantize gravity while strictly preserving the total Hamiltonian zero necessitates an approach that realizes background freedom, namely, background independence in quantum spacetime.

To begin with, the definition of particles is contingent upon the existence of a fixed background spacetime, typically assuming flat spacetime where the Riemann tensor vanishes. This is nothing more than a background-dependent method that treats the spacetime as absolute. Gravitons are merely particles moving on the background spacetime that are valid only when their presence does not affect the spacetime structure. This assumption may hold true in our immediate world, but it is by no means valid in the trans-Planckian domains involving extreme gravity, such as the central region of a black hole.

In this article, through arguments grounded on the fundamental principles of gravity and quantum field theory, we challenge the prevailing consensus on the ghost issue shared by many researchers. The argument therein reveals an uncharted aspect of the early universe and the interior of a black hole.

\section{Total energy-momentum tensor identically vanishes}

Because the gravitational field ontologically constitutes spacetime itself, it plays a fundamentally distinct role from matter fields defined under the existence of spacetime. The gravitational field does not presuppose any absolute time and space, and treats the spacetime structure itself as a variable that changes dynamically depending on matter and energy. This property inherent to gravity is referred to as background independence in a broad sense, as a thought beyond Newtonian and special relativistic predecessors relying on a fixed spacetime to serve as the stage for physical phenomena \cite{vassallo}.

The primary step in the quantization of the gravitational field is to ensure that the path integral can be consistently evaluated in an ultraviolet (UV) complete manner. In this endeavor, introducing a finite UV cutoff, which is equivalent to discretizing spacetime, should be avoided, because it leads to problems such as the breaking of diffeomorphism invariance, the loss of renormalizability, and the cosmological constant problem \cite{hamada22, springer}. Preserving diffeomorphism invariance is postulated as the most critical guiding principle, which dictates the field continuity, consequently imposes the stringent condition of renormalizability upon the gravitational path integral.

From the 1960s to the early 1980s, the necessity of higher-derivative gravitational actions was argued as a clue to solving the problem of renormalizability \cite{ud, veltman, stelle, tomboulis, ft}. However, the emergence of the ghost problem posed a significant challenge,\footnote{The argument presented in this article is different from traditional ones based on the idea of Lee and Wick \cite{lw}, rather closer to the perspective of Boulware, Horowitz, and Strominger \cite{bhs}.} leading to a decline in research on quantum gravity based on quantum field theory. The quantization method employed at that time was the weak-field approximation, which treats gravitons as quanta. We argue that the failure of this approach is intrinsic to the approximation itself rather than the underlying action.

First of all, we will examine the fundamental constraints that invariably emerge when formulating classical and quantum theories of gravity. Conventionally, the classical Einstein equation is presented as $R_{\mu\nu} - \half g_{\mu\nu} = 8\pi G \, T^{\rm M}_{\mu\nu}$  with matter energy-momentum tensor $T^{\rm M}_{\mu\nu}$ placed on the right-hand side as an external source driving spacetime curvature. This aligns with Mach's principle, which states that the structure of spacetime is determined by the matter present within it.  However, this description obscures the essence of diffeomorphism invariance.

The gravitational field universally couples to all fields, and the quantity defined by the functional variation of the action $I$ with respect to the gravitational field, namely $T^{\mu\nu} = \fr{2}{\sq{-g}} \fr{\dl I}{\dl g_{\mu\nu}}$, serves as the rigorous definition of the energy-momentum tensor. From this perspective, the Einstein equation should be expressed in the vanishing of the total energy-momentum tensor as $T_{\mu\nu}= M_{\rm P}^2 \, ( - R_{\mu\nu} + \half g_{\mu\nu} R ) + T^{\rm M}_{\mu\nu} = 0$, where $M_{\rm P} =1/\sq{8\pi G}$ is the reduced Planck mass. In particular, the time-time component of the energy-momentum tensor represents the Hamiltonian density, which yields the Hamiltonian when integrated over the spatial volume.  Consequently, the equation of motion for the gravitational field states that the total Hamiltonian of the system vanishes.

In quantum field theory, equations of motion manifest as functional  identities known as Schwinger-Dyson equations. Employing the path integral formalism, for a generic field $f$ with an action $I[f]$, this identity is expressed as $\int [df] \fr{\dl}{\dl f(x)} e^{iI[f]} = i \lang \fr{\dl I[f]}{\dl f(x)} \rang = 0$, where a spacetime manifold without boundaries is considered. In this manner, when incorporating quantum theory, the variational principle is inherently derived as an identity.

Now, we apply the Schwinger-Dyson equation to the gravitational field. Here, we discuss this equation in a general framework without specifying the explicit form of the gravitational action while assuming the field continuity or renormalizability. Since varying the action with respect to the gravitational field yields the energy-momentum tensor, we obtain an identity such that its expectation value vanishes as follows:\footnote{To be precise, the path integral measure has to be defined properly involving the Wess-Zumino actions for conformal anomalies \cite{springer, riegert}, but this form holds true if dimensional regularization is employed \cite{hamada02, hamada14}.}
\begin{eqnarray*}
    \int [dg] \fr{\dl}{\dl g_{\mu\nu}(x)} e^{i I[g]} = \fr{i}{2} \lang \sq{-g} \, T^{\mu\nu}(x) \rang  = 0 .
\end{eqnarray*}

This identity constitutes a fundamental equation of quantum gravity, corresponding to what are commonly called the Hamiltonian and momentum constraints. For simplicity, explicit matter field contributions have been omitted here, but since matter fields are intrinsically incorporated into the action $I$, $T^{\mu\nu}$ represents the energy-momentum tensor of the entire system including matter fields. Here it is critical to note that while the equation of motion and the energy-momentum tensor are distinct entities in conventional fields, they are intrinsically the same in the gravitational field.

Thus, whether classical or quantum, the gravitational equation of motion demands the vanishing of the total energy-momentum tensor. This equation represents background independence in the sense that spacetime is not predetermined but rather derived as a result of solving the equation. In quantum gravity, it is elevated to a property that the theory holds true independent of any background spacetime selected as a reference.

This constitutes a fundamental departure from the architecture of standard quantum field theory. In standard frameworks, quantum field theory is formulated around Hamiltonian eigenstates defined on a background spacetime with absolute significance, and then a state with a zero eigenvalue represents the vacuum, and states with positive eigenvalues are identified as physical particle excitations.

This leads to an immediate corollary: if all eigenvalues are positive, the total Hamiltonian is never zero. This implies that if quantum gravity is treated within the standard framework of quantum field theory, only the vacuum state is permissible, leaving nothing behind. In other words, it indicates that negative energy is required to define gravitational systems.

Herein lies the crux of the quantum gravity problem. To reiterate, the constraint wherein the total Hamiltonian vanishes emerges as an intrinsic consequence of quantizing gravity. To reconcile gravity with quantum theory, we should rigorously interpret the physical implications of this constraint. This implicitly suggests that formulating quantum gravity within the conventional framework of Lorentz-invariant quantum field theory will result in failure. Ultimately, quantum gravity is not merely a discipline dedicated to graviton scattering amplitudes, but rather a profound inquiry into the fundamental nature and dynamical structure of spacetime itself.

\section{Gravitational systems are exempt from Ostrogradsky's theorem}

At this juncture, it is imperative to note the profound relevance of Ostrogradsky's theorem \cite{ostrogradsky, pu, woodard, gn}, established in 1850, to the aforementioned discussion.

This classic theorem underpins the fact that, since the inception of Newtonian mechanics, physical equations of motion have conventionally been formulated using second time derivatives at most. The theorem formally demonstrates that in dynamical systems governed by equations of motion with higher time derivatives, the Hamiltonian becomes unbounded from below, thereby rendering the system unstable with a bottomless energy spectrum. In theoretical physics, generic modes that obstruct the boundedness of energy are designated as ``ghost modes". Therefore, the theorem asserts that dynamical systems featuring higher time-derivative terms are inherently unphysical due to the inevitable presence of these unphysical ghost modes.

The motivation for formulating the Einstein equation using functions restricted to second derivatives stems not only from the precedent set by Newtonian equations but also from the existence of this theorem. However, despite being formulated strictly with second derivatives to circumvent Ostrogradsky's theorem, the Einstein equation inherently involves a ghost mode. Its existence originates from the fact that the Einstein-Hilbert action is constructed by the scalar curvature, which takes values across the entire real domain ($-\infty < R < \infty$). Thus, the action is not bounded from below, indicating the absence of a stable energy ground state. The gravitational mode responsible for this unboundedness, the conformal mode in this case, acts as a ghost mode, thereby causing instability.

Nevertheless, this ghost mode constitutes a physically indispensable component. Indeed, the Friedmann equation $-3 M_{\rm P}^2 H^2 + \rho =0$ is rendered untenable without the ghost mode producing negative energy contribution $-3 M_{\rm P}^2 H^2$, where $H$ denotes the Hubble variable originating from the conformal mode of the gravitational field while $\rho$ is a positive energy density of matter. Furthermore, the large-scale structure of the current universe arises from the gravitational amplification of primordial fluctuations, which is driven by the presence of this unstable mode.

The ghost mode in the gravitational field is identified as a constrained ghost mode that exists under the condition of a vanishing total Hamiltonian. Therefore, this mode is fundamentally distinct from freely propagating ghost modes addressed by Ostrogradsky's theorem. While free ghost modes are physically prohibited,  the constrained gravitational ghost modes implicitly underpin the structure of spacetime.

Hitherto, although we have not addressed the specific action of quantum gravity, the vanishing of the Hamiltonian is universally guaranteed, provided the path integral is rigorously formulated to preserve diffeomorphism invariance. In stark contrast, the crux of Ostrogradsky's theorem is that higher-derivative theories inherently possess ghost modes, rendering the Hamiltonian unbounded from below.

Hence, Ostrogradsky's theorem fundamentally fails to apply to gravitational systems where the issue of unboundedness is circumvented by the Hamiltonian constraint dictating a vanishing total Hamiltonian. Accordingly, the gravitational action should be determined free from this theorem. Rather, irrespective of whether we consider classical or quantum gravity, the presence of ghost modes is essential to ensure that the Hamiltonian vanishes identically.

In this context, it is worthwhile to clarify theoretical limitations generally encountered when adopting the weak-field approximation, in which the gravitational field is expanded around a specific background spacetime. We will explore this framework below to systematically elucidate the underlying causes of the ghost problem.

In the weak-field approximation, a solution to the gravitational equations of motion, typically the Einstein equation, serves as the background spacetime. Within this framework, all particles, including gravitons, are formulated as Hamiltonian eigenstates with positive eigenvalues defined on this background. If flat spacetime is chosen as the background, this approach reduces to a standard Lorentz-invariant quantum field theory.

Since the Hamiltonian can never be zero in a physical system composed of particles with positive energy, this approach implicitly assumes that the total Hamiltonian can be considered as approximately zero. This implies that the characteristic particle energy is negligibly small compared to the Planck energy. Therefore, this methodology is solely applicable in sufficiently low-energy regime, where spacetime distortions induced by particles, or back-reactions, remain negligible, so that particles allow to be treated as point-like entities.

It should be noted that while requiring all particles to possess positive energy appears to contradict the equations of motion for gravity, it is a prerequisite for the validity of the weak-field approximation. This is because the presence of an unconstrained free ghost mode would cause its energy to decrease indefinitely, deviating unboundedly from zero. This invalidates the premise of the approximation, which assumes that the Hamiltonian remains close to zero. Thus, adopting the weak-field approximation inevitably renders Ostrogradsky's theorem valid, implying that this framework is only applicable to gravitational theories in which dominant kinetic terms are described by at most second derivatives.

In fact, applying the weak-field approximation to Einstein's theory of gravity enables the elimination of unphysical ghost modes, yielding a Lorentz-invariant system of physical gravitons with positive energy. This accounts for the widespread adoption of this approach. This framework is completely ghost-free, but when formulating the theory in a curved background, ghost modes contribute only as structural elements of the background geometry.

\section{Gravitational actions for trans-Planckian physics}

Returning to the issue of the gravitational action, it is worth noting that since the gravitational field is inherently dimensionless, fourth-derivative gravitational actions naturally contribute as dimensionless quantities in the UV limit. Moreover, these actions render the theory renormalizable.

Incorporating the Riemann tensor directly overcomes the drawback that Einstein equation lacks this fundamental tensor that governs the field strength of gravity. Consequently, introducing positive-definite quadratic terms of the Riemann tensor not only renders the theory renormalizable but also naturally resolves spacetime singularities, because from the perspective of quantum field theory, singular solutions yielding divergent actions are unphysical and thus discarded. Then, we can conclude that no singularities exist inside a black hole, and that it is obliterated by quantum fluctuations \cite{hamada20}.

Here it should be noted that while the ghost mode in Einstein's theory of gravity arises from the indefiniteness of its action, ghost modes in fourth-derivative gravitational theories exist despite their actions being positive-definite, thus ensuring stability.

In the trans-Planckian world, fourth-derivative terms become dominant, and there it is imperative to establish a quantization scheme wherein the total energy-momentum tensor vanishes, rendering Ostrogradsky's theorem inapplicable. This essentially entails quantization of spacetime itself, addressing a framework characterized by intense quantum spacetime fluctuations.

In quantum spacetime where the gravitational field undergoes violent quantum fluctuations, the classical notion of time and space breaks down. Thus, even when flat spacetime is adopted as a reference background, prominent fluctuations around it render such a choice devoid of physical significance. This implies that physical distances become virtually unmeasurable due to these fluctuations. This property, recognized as background freedom or background independence in a quantum sense, epitomizes a truly scale-invariant world, where not only do dimensionful physical constants fail to contribute to the dynamics, but distance itself ceases to make sense. This very featureless state is ideal for the beginning of the universe.

The theoretical exploration of the trans-Planckian physics necessitates the implementation of non-perturbative frameworks. In this direction, a promising candidate is a renormalizable and asymptotically background-free quantum gravity formulated based on a certain conformal field theory with positive-definite conformally invariant fourth-derivative gravitational actions, in which the intense fluctuations in distance are expressed by treating the conformal mode of the gravitational field with particular rigor \cite{springer}.

In this theory, the background freedom is represented as a special conformal symmetry that naturally emerges as an intrinsic feature of diffeomorphism invariance in the UV limit. That is called BRST conformal invariance, which implies that all of worlds with different scales connected by conformal transformations become gauge-equivalent. The BRST condition for this conformal invariance represents nothing other than the Hamiltonian and momentum constraints. Under this condition, all ghost modes in the fourth-derivative gravitational field become unphysical and physical states are given solely by composite scalar quantities, without any tensor ones \cite{hamada12}. This result is consistent with current cosmological observations \cite{hy, hamada24a}.

\section{Outlook}

The flow of time emerges dynamically through changes in the state, subject to the constraint that the total Hamiltonian vanishes. While the uniform progression of time holds no meaning in a background-free quantum gravity framework, an approximate notion of time was established, manifesting as a monotonic increase of the conformal mode, through structural changes comprising cosmic inflation and the subsequent spacetime phase transition. During the inflationary period, large spacetime fluctuations of order $\dl R/R \sim O(1)$ were dynamically attenuated to the level of $O(10^{-5})$ \cite{hamada24a}, giving rise to the current homogeneous and isotropic classical universe and to time as a persistent physical reality. Crucially, this emergence of time is made possible only by the existence of ghost modes inherent to the gravitational field.

In a universe lacking absolute time, conventional conservation laws that depend on time coordinates do not apply. Here, a conservation law signifies that a physical quantity remains invariant throughout the evolution of the universe and is formulated as a renormalization-group invariant \cite{hamada22, collins, bc, hamada24b}. In the framework of quantum gravity, the strict vanishing of the energy-momentum tensor ensures the exact conservation of energy and momentum.

In his monograph, {\it The Inflationary Universe: The Quest for a New Theory of Cosmic Origins}, Guth characterizes the emergence of a matter-filled universe from a state of zero total energy as the ``ultimate free lunch" \cite{guth}. He draws this analogy because matter appears to be generated from nothing without any net physical energy cost. Broadly, any gravitational system where the positive energy of matter and the negative energy of gravity exactly offset each other can be considered  in a free lunch state. The universe has always existed, never suddenly manifest ex nihilo; undergoing continuous changes in its state while perpetually preserving a total energy of zero.

The existence of ghost modes is an indispensable prerequisite for formulating the gravitational field dynamics wherein the total Hamiltonian vanishes. The cosmic expansion is exactly caused by ghosts. However, we do not observe the ghosts directly; rather, we recognize this fact by observing celestial bodies receding in all directions. In the conceptual sense of being physically unobservable yet dynamically operative, these entities are reminiscent of Bohmian hidden variables \cite{gilder}. Nevertheless, within Lorentz-invariant quantum systems, the violation of Bell's inequalities, demonstrated by Aspect, Clauser, and Zeilinger, rigorously establishes the nonexistence of local hidden variables. Ultimately, ghosts can be formally understood as hidden degrees of freedom that are activated exclusively in physical regimes where gravitational dynamics are essential.


\end{document}